\documentclass[aps,prb,twocolumn,groupedaddress,showpacs,showkeys]{revtex4}
\usepackage[dvips]{graphicx,color}

\newcommand{\fatq}{\mathbf{q}}

\newcommand{\micron}{$\mu$m}

\newcommand{\pgnfigure}[2]{\begin{figure}\includegraphics[height=6cm,clip=true]
{#1.eps}\caption{\label{#1}#2}\end{figure}}
\newcommand{\pgnfigurestwo}[3]{\begin{figure}\includegraphics[height=6cm,clip=true]
{#1.eps} \par \par\includegraphics[height=6cm,clip=true]{#2.eps}
\caption{\label{#1}#3}\end{figure}}
\newcommand{\pgnfiguresthree}[4]{\begin{figure}\includegraphics[height=5.5cm,clip=true]
{#1.eps} \par \par
\includegraphics[height=5.5cm,clip=true]{#2.eps} \par \par
\includegraphics[height=5.5cm,clip=true]{#3.eps}
\caption{\label{#1}#4}\end{figure}}

\begin{document}


\title{Field-induced non-Fermi-liquid resistivity\\
of stoichiometric YbAgGe single crystals}



\author{P. G. Niklowitz} \email[e-mail:]{niklowit@drfmc.ceng.cea.fr}
\author{G. Knebel}
\author{J. Flouquet}
\affiliation{D\'{e}partement de Recherche Fondamentale sur la
Mati\`{e}re Condens\'{e}e, SPSMS, CEA Grenoble, 38054 Grenoble
Cedex 9, France}
\author{S. L. Bud'ko}
\author{P. C. Canfield} \affiliation{Ames
Laboratory, Iowa State University, Ames, Iowa 50011, USA}
\affiliation{Department of Physics and Astronomy, Iowa State
University, Ames, Iowa 50011, USA}


\date{\today}

\begin{abstract}

We have investigated hexagonal YbAgGe down to 70~mK by measuring
the magnetic-field and temperature dependence of the resistivity
$\rho$ of single crystals in fields up to 14~T. Our results extend
the $H-T$ phase diagram to the lowest temperatures for $H$ applied
in the basal plane and along the $c$-axis. In particular, critical
fields for the suppression of several magnetic phases are
determined. The temperature dependence of $\rho(T)$ is unusual:
whereas at low $H$, $\rho(T)$ reveals a temperature exponent $n\ge
2$, we find $1\le n<1.5$ and strong enhancement of the temperature
dependence of $\rho(T)$ close to and beyond the highest critical
field for each field direction. For $H$ applied in the basal
plane, at high fields a conventional $T^2$ dependence of $\rho(T)$
is reached above 10~T accompanied by an approach to saturation of
a strong drop in the residual resistivity. YbAgGe appears to be
one of few Yb-based stoichiometric systems, where quantum-critical
behaviour may be induced by a magnetic field.

\end{abstract}


\maketitle

\section{Introduction}

Metals at the border of magnetic order often show deviations from
the thermodynamic and transport properties predicted by
Fermi-liquid (FL) theory, the standard model of metals. Materials
can be tuned through the magnetic quantum phase transition by
application of pressure or doping. A different route lies in
tuning materials by the application of a magnetic field.
Field-induced quantum phase transitions have been studied in
several groups of systems: (i) the extensively studied materials
with a metamagnetic transition
CeRu$_2$Si$_2$,\cite{hae87a,flo02a}\
Sr$_3$Ru$_2$O$_7$,\cite{gri01a}\ UPt$_3$,\cite{kim00a}\ or
URu$_2$Si$_2$;\cite{kim03a}\ (ii) Ce$_2$IrIn$_8$, where
non-Fermi-liquid behaviour might be linked to field-induced
magnetic order other than ferromagnetic order;\cite{kim04a}\ (iii)
CeCoIn$_5$, where non-Fermi-liquid behaviour has been found at the
upper critical field of the superconducting phase, below which
superconductivity and antiferromagnetism seem to
coexist;\cite{pag03a}\ (iv) antiferromagnetic metals where
spontaneous magnetic order gets suppressed. Here we focus on group
(iv). Field-induced quantum critical points (QCPs) have already
been identified in several Ce-based antiferromagnetic
heavy-fermion systems, including CeCu$_{6-x}$Ag$_x$ or
CeCu$_{5.8}$Au$_{0.2}$.\cite{heu98a,sch99b,ste01b}\ Among Yb-based
stoichiometric heavy-fermion systems, only YbRh$_2$Si$_2$ has been
reported to show field-induced quantum-critical
behaviour.\cite{geg02a,ish02a,cus03a}

YbAgGe offers a new possibility for a study of the rich f-electron
physics at a field induced magnetic quantum phase transition in a
stoichiometric compound. YbAgGe has recently been recognised as a
new heavy-fermion system with a linear specific heat coefficient
$\gamma$ of a few hundred mJ/molK$^2$ at low temperatures and a
Kondo temperature $T_K\approx 25$~K.\cite{mor04a,kat04a}\ Two
antiferromagnetic transitions, which are found at low temperatures
already at ambient conditions, can be fully suppressed within the
experimentally well accessible magnetic field regime of less than
10~T.\cite{bud04a}

YbAgGe orders in the hexagonal ZrNiAl-type
structure.\cite{mor04a}\ The Yb$^{3+}$ ions are exposed to a
crystal field with an orthorhombic point symmetry, which splits
their eight-fold $J=7/2$ multiplet into four doublets. Inelastic
neutron scattering revealed a crystal-field excitation at 12~meV,
which is also visible as a Schottky anomaly in the specific heat
at 60~K.\cite{mat04a,kat04a}\ The specific heat data can be
explained, if it is assumed that the Schottky anomaly corresponds
to the lowest crystal-field excitation. However, such a model does
not explain the susceptibility and some uncertainty on the
crystal-field scheme remains. Magnetisation measurements show
clear anisotropy growing to $\chi_{ab}/\chi_c\approx 3$ at low
temperatures.\cite{mor04a,kat04a}\ Saturation of the magnetisation
close to 15~T is only found when the field is applied in the easy
plane. Above $T_K$ the susceptibility derived from the
magnetisation data follows the Curie-Weiss law with an effective
moment of $4.4\mu_B$ (close to the free-ion value for Yb$^{3+}$ of
$4.5\mu_B$) and with a Weiss temperature $\Theta=-30$~K, which
suggests antiferromagnetic interactions between the moments. Below
$T_K$ the susceptibility levels off and shows a weak maximum at
about 4~K\cite{mor04a,kat04a}.

At low temperatures, two transitions to antiferromagnetic phases
(AF1 at the lowest temperatures and AF2) are observed at zero
field.\cite{mor04a,bud04a,ume04a,fak04a,fak05a}\ A first
transition is located at $T_1=0.65$~K, which has been observed as
a sharp peak in the heat capacity and as a jump in the
resistivity. A clear hysteresis in the heat capacity and in the
resistivity shows that the transition is first
order.\cite{mor04a,bud04a,ume04a}\ At $T_2=0.9\pm0.1$~K there is a
second transition, which shows up as a relatively broad feature in
the specific heat and as a broad change of slope in the electrical
resistivity. Further information on the nature of the
antiferromagnetic phases comes from heat capacity measurements and
neutron scattering studies. The entropy calculated from the heat
capacity reaches only 5\%\ of $R\ln 2$ at 1~K and $R\ln 2$ close
to 25~K, which suggests that any ordered moment is only of the
order of 0.1~$\mu_B$.\cite{mor04a}\ Recently, neutron diffraction
at zero field revealed for the AF1 phase the commensurate ordering
wave vector (1/3,0,1/3) and indicated moment orientation
predominantly in the basal plane.\cite{fak04a}\ The AF2 phase
above $T_1$ shows incommensurate order and its suppression at
$T_2$ happens via a second-order transition.\cite{fak05a}\
Inelastic neutron scattering (Ref.~\onlinecite{fak04a}) was found
to be quasielastic and revealed fluctuations predominantly in the
basal plane with a temperature dependence characteristic of a
heavy-fermion material. At low temperatures, however, an anomalous
$\fatq$-dependence of the linewidth $\Gamma(\fatq)$ was found,
which varies along the $c$-axis but which is constant in the basal
plane. The origin of this anomalous $\fatq$-dependence might come
from the structural particularity of YbAgGe that the magnetic
Yb-ions lie on a distorted Kagome lattice with the potential for
magnetic frustration.

YbAgGe completes a trend, which can be observed in RAgGe
compounds, where R stands for various rare earths.\cite{mor04a}\
In going from R=Tb to R=Yb the magnetic anisotropy changes from
being axial $\chi_{ab}/\chi_c<1$ to being planar
$\chi_{ab}/\chi_c>1$. Furthermore, the magnetic transition
temperatures decrease with an approximate de-Gennes scaling.
However, in YbAgGe, the large value for the linear heat-capacity
coefficient $\gamma$ indicates that magnetic exchange connected
with the hybridisation between f-electrons and conduction
electrons influences the value for the magnetic transition
temperatures and it is by accident that YbAgGe does not
significantly deviate from de-Gennes scaling of the RAgGe series.
The large $\gamma$ value also suggests that YbAgGe should be close
to quantum phase transitions and possibly a quantum critical
point.

The proximity of quantum phase transitions has already been
pointed out by the exploration of the $H^a$-$T$ and $H^c$-$T$
phase diagram ($a$ and $c$ indicate the crystal axis, along which
the magnetic field was applied).\cite{bud04a,ume04a,fak05a}\ In
addition to the above mentioned signals in the heat capacity and
resistivity indications of the two transitions are also visible in
the magnetoresistance and in the magnetization. For both field
directions applied magnetic fields of less than 10 T are
sufficient to reach the critical fields $H_c$ necessary to
suppress antiferromagntic magnetic order. From neutron scattering
experiments $H^a_{c2}$ ($H_{c2}$ for $H\|a$) was found to be
3~T.\cite{fak05a}\ However, features in the resistivity, heat
capacity, and magnetization\cite{ume04a}\ extrapolate to a further
critical field $H^a_{c3}$. Furthermore, the critical field
$H^a_{c3}$ is the starting point of the lines in the $H-T$ phase
diagrams, which is defined by features in the Hall resistivity,
and which shifts to higher field with increasing
temperature.\cite{bud04b}\

This paper describes detailed measurements of low-temperature
properties of YbAgGe in a magnetic field. The measurements cover
the field range up to 14~T with the field applied in the easy
plane and along the c-axis. The current has been applied along two
crystallographic directions as well. The corresponding $H^a-T$ and
$H^c-T$ phase diagrams are extended to 70~mK, which allows a
precise determination of the critical fields. One specific aim was
to search for field-induced quantum critical phenomena. Specific
heat and resistivity measurements down to 0.4~K indicated the
existence of quantum critical fluctuations: in approaching
$H^a_{c3}$, $\gamma$ and the strength of the low-temperature
dependence of the resistivity are enhanced and the resistivity
deviates more strongly from the quadratic Fermi-liquid
form.\cite{bud04a}\ This low-temperature study shows quite
dramatic changes in the temperature dependence of the resistivity
just above these critical fields, which are typically observed
close to a quantum critical point. However, an unconventional form
of the resistivity has been detected over unusually large
$H$-intervals down to very low temperatures. Additionally, a
strong field dependence of the residual resistivity with a large
enhancement at low fields has been observed.

\section{Experimental}

The YbAgGe samples measured in this study were grown from
high-temperature ternary solutions rich in Ag and Ge. The samples
had the form of clean hexagonal cross-section rods of several
millimeters length and 0.3-0.8~mm$^2$ cross section. Their
structure and the absence of impurity phases were confirmed by
powder X-ray diffraction (details of the sample growth are found
in Ref. \onlinecite{mor04a}). The ac-resistance was measured by a
standard four-terminal method ($f\approx 28$~Hz,
$I=25-100$~$\mu$A). Gold wires were attached to the samples by
spot-welding using the smallest possible voltages for welding. The
samples were cleaned in HNO$_3$ before attaching the gold wires to
remove residual flux and in general to avoid damage to the sample
during spot-welding. For measurements with the current in the
plane thin slices with a thickness of 80-300~\micron\ were cut off
the sample rods. For the in-plane measurements the current was
approximately sent along the [100] direction. $\rho(T,H)$ was
measured down to 70~mK in a dilution refrigerator containing an
Oxford Instruments 18~T superconducting magnet. The sample was
orientated such that the field was either applied along [001] or
in the plane approximately along [100]. 
A low-temperature RuO$_2$ thermometer was initially only
calibrated in zero field. The in-field calibration was done by
regulating a capacitor to a constant value during field sweeps.
The optimal balance in the calibration procedure between
eddy-current heating and a drift of the capacitance resulted in a
temperature error of 3~mK for the RuO$_2$ thermometer. This is the
dominant error in the resistivity measurements and determines the
errors of temperature exponents in the analysis of the resistivity
data. The resistivity was determined in temperature sweeps with a
sweep rate of 0.2~K/h and in field sweeps with a sweep rate of
3~T/h. Due to the hysteretic character of the transition $T_1(H)$
we specify here that in the hysteretic region results have been
gained from up-sweeps in field and temperature unless otherwise
indicated. Furthermore, before each up-sweep in temperature the
sample was zero-field cooled.

\section{Results}

\pgnfigure{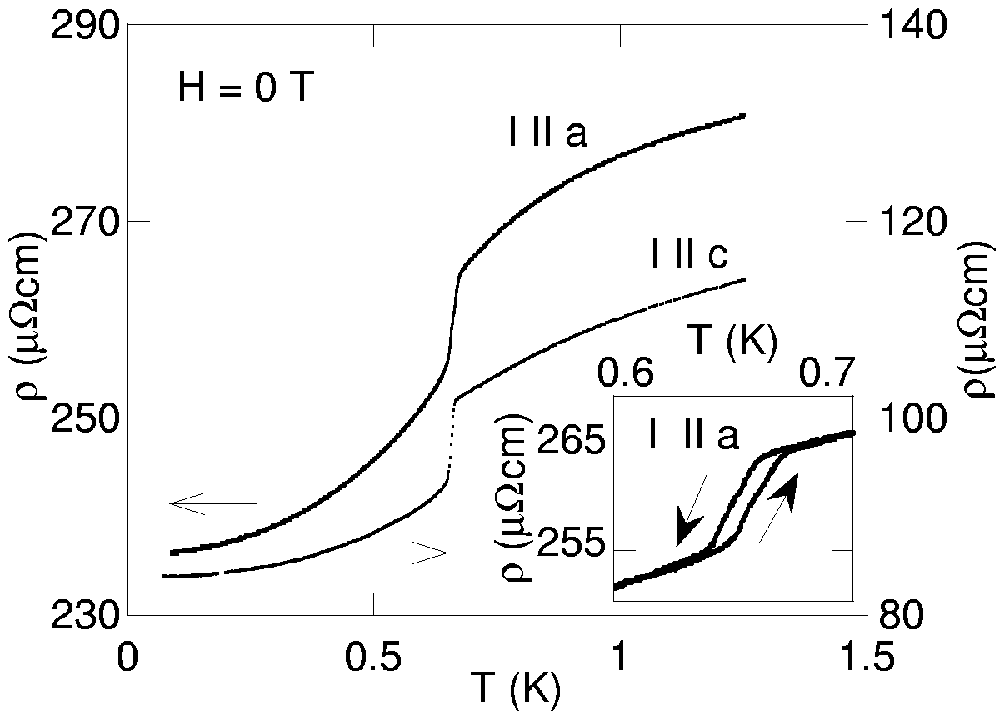}{Temperature dependence of the
resistivity with no applied field. No qualitative change in the mK
range is found upon altering the current direction. However, the
resistivity is anisotropic with larger in-plane values. In the
inset the first order character of the transition between the AF1
and AF2 phase is shown. Arrows indicate the direction of the
temperature sweeps.}

In Figure~\ref{YbAgGemtswp1tr}, we show the resistivity of YbAgGe
at zero field. For the in-plane as well as for the $c$-axis
resistivity we find a large jump of the order of 10~$\mu\Omega$cm
at $T_1(H=0\mbox{T})=666\pm 5$~mK as signature of the transition
between the AF1 and AF2 phase. The hysteresis in the inset of
Figure~\ref{YbAgGemtswp1tr}\ confirms the first-order character of
this transition. No sharp signal for $T_2$ of the second
transition at around 1~K can be resolved. A broad change of slope
of $\rho(T)$ around $T_2$ is more evident in previous studies,
which reached to higher temperatures.\cite{mor04a,bud04a,ume04a}\
The in-plane resistivity is in general higher and shows a stronger
temperature dependence. The difference might partially arise from
the error in the estimation of the small sample geometry. For the
resistance ratio (RRR) we find 2.7 and 3.4 for the current
directions $I\|a$ and $I\|c$, respectively. The residual
resistivity $\rho_0$ is 2.8~times higher in the plane than along
the c-axis. The order of magnitude of $\rho_0$ in our measurements
is comparable to values found in previous zero-field studies.
Values for $T_1(H=0)$ from previous studies, ranging from 0.55~K
to 0.65~K, are close to our result as
well.\cite{mor04a,kat04a,bud04a,mat04a,ume04a}\ However, we cannot
confirm the previously found sudden rise of the in-plane
resistivity towards lower temperatures at $T_1$.\cite{ume04a}\
Differences in the sample growth technique might explain the
discrepancies.

In the following we present the resistivity of YbAgGe in a
magnetic field. We measured the resistivity for the combinations
of field and current direction $H\|a$ with $I\|a$ or $I\|c$, and
$H\|c$ with $I\|c$. Since it will be seen that for $H\|a$ the
temperature dependence $\Delta\rho=\rho-\rho_0$ is rather
independent of the current direction we discuss the results for
the two current directions in parallel. For the same reason we
have done only measurements with one current direction in the case
of $H\|c$. The resistivity curves $\rho(T)$ have been analysed by
fitting $\rho(T)=\rho_0+cT^n$ ($\rho_0$, $c$, and $n$ are fitting
parameters) over the largest temperature interval, which allows a
single-power-law fit. This analysis is used to determine the
temperature-independent residual resistivity $\rho_0$, and
separate it from the temperature dependent part $\Delta\rho=cT^n$,
where $c$ is a constant and $n$ the temperature exponent of the
resistivity.

\subsection{$H\|a$}

\pgnfiguresthree{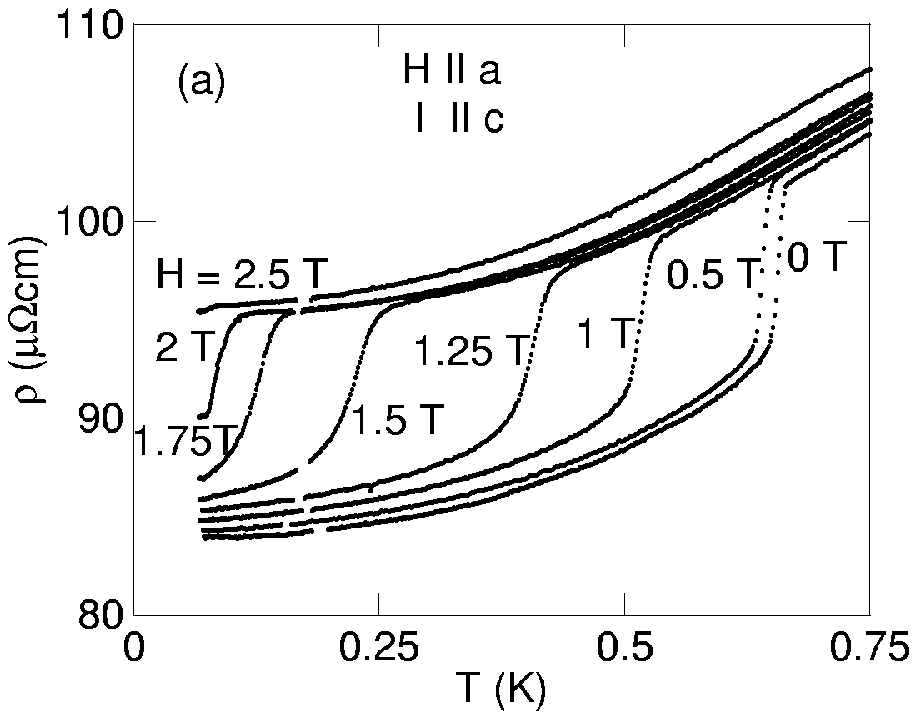}{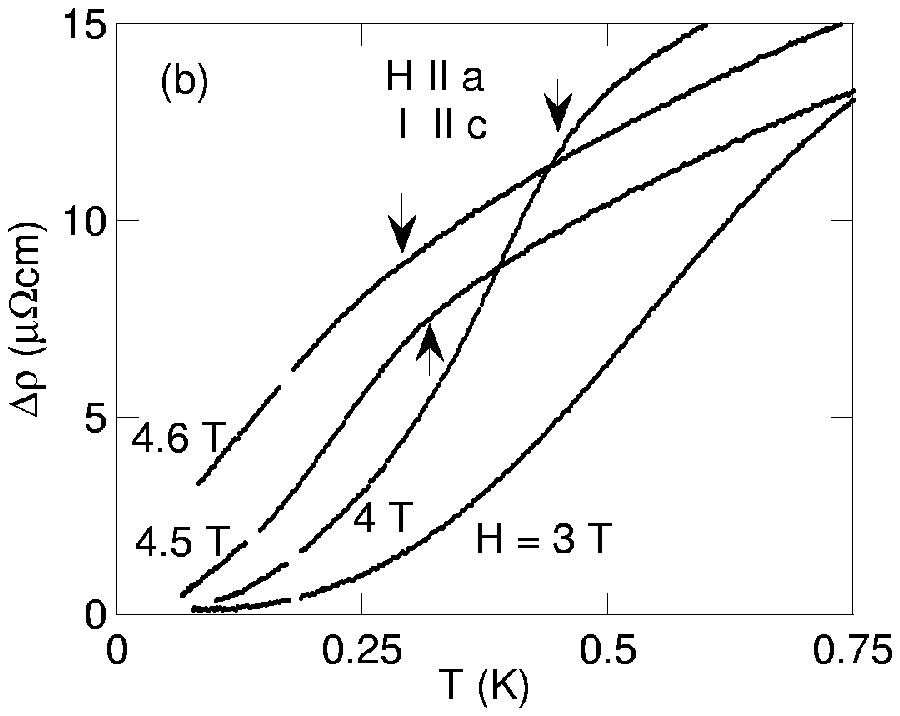}{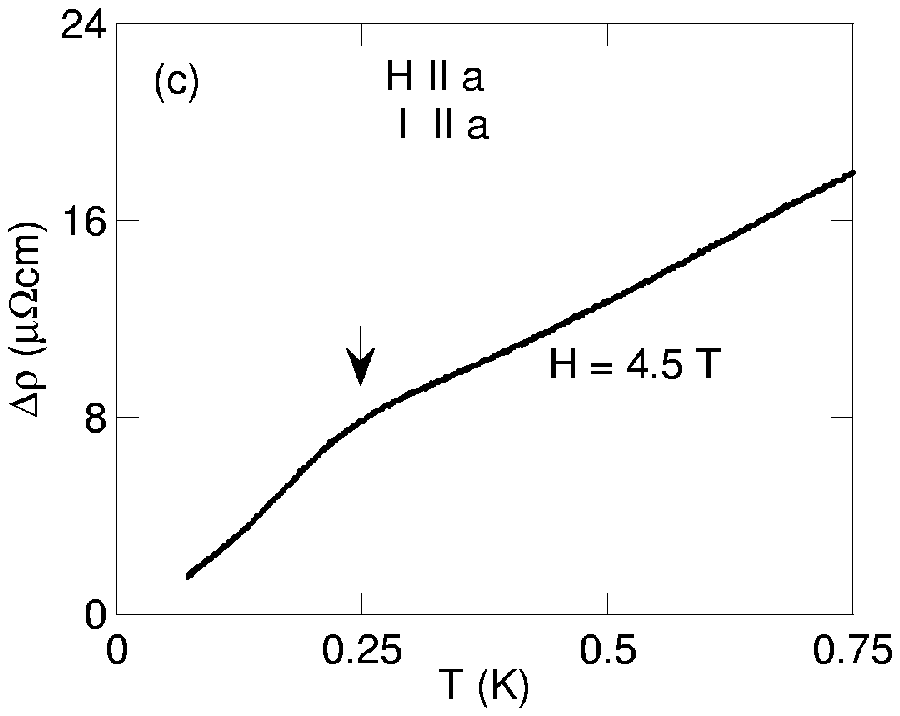}
{Signatures in the temperature dependence of the resistivity
$\rho(T)$ for $H\|a$. (a) $T_1$ (onset of drop in $\rho$) is
gradually suppressed with field and fully disappears at 2.5~T.
(b),(c) Signatures of $T_3$: the shoulder (marked by arrows) also
is suppressed with increasing field. In parallel, the initial
slope of $\Delta\rho(T)$ increases from 3 to 4.6~T.}

\pgnfigure{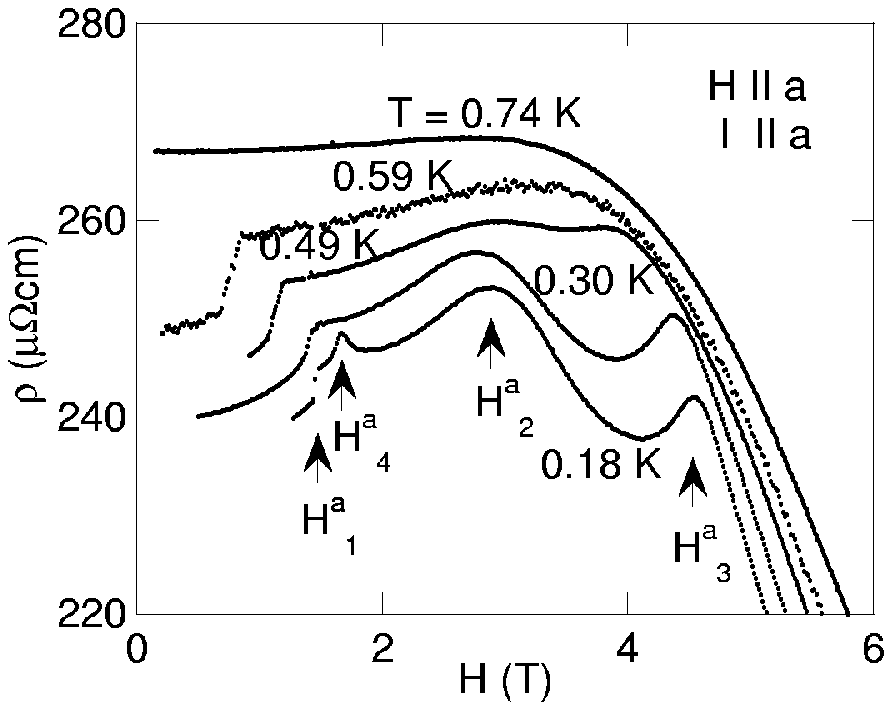}{Signatures in the field dependence of
$\rho$ for $\rho(H\|a)$ at various temperatures. The field sweeps
at different temperatures reveal the first-order transition
($H^a_1$, onset of jump in $\rho$), the second-order transition
($H^a_2$, bump) and a further feature ($H^a_3$, onset of strong
drop of $\rho$ towards high fields), which indicates a further
transition. A forth feature ($H^a_4$), has yet to be confirmed by
further measurements.}

In this section we present results of the temperature and field
dependence of the resistivity for the case $H\|a$.
Figure~\ref{YbAgGemcatswp1tr}a contains $\rho(T)$ curves at small
fields showing signatures of the first-order transition between
the AF1 and AF2 phase and their suppression to low temperatures
towards higher fields. Up to its suppression below 70~mK at 2.5~T
the first-order transition is marked by the onset of a large drop
of $\rho(T)$ at $T_1(H\|a)$, similar to its transition signal at
zero field. Figure~\ref{YbAgGemcatswp1tr}b shows a much less
visible feature in $\rho(T)$ that manifests itself only as a weak
shoulder at $T_3(H\|a)$. This shoulder in $\rho(T)$ can be
identified up to 4.6~T. Signatures corresponding to $T_1(H\|a)$
and $T_3(H\|a)$ are clearly seen in the magnetoresistance
(Figure~\ref{YbAgGemaahswptr}). The first-order transition appears
as the onset of a drop towards low field of $\rho(H\|a)$ at
$H^a_1(T)$, similar to its feature in $\rho(T)$. The feature at
$T_3(H\|a)$ shows up as a shoulder at $H^a_3(T)$ and is better
visible at lower temperatures. A broad bump indicates a further
transition at $H^a_2(T)$, which can be followed up to
$\approx$500~mK. At low $T$ there might be a forth feature close
to $H^a_1(T)$, which has to be confirmed by even more detailed
magnetoresistance measurements at low temperatures. The signatures
of transitions in $\rho(T,H^a)$ shown in
Figures~\ref{YbAgGemcatswp1tr}\ and \ref{YbAgGemaahswptr} and
further results of $\rho(T,H^a)$ measurements, which are not
explicitly shown, allow us to draw the $H^a$-$T$ phase diagram of
YbAgGe (Figure~\ref{YbAgGetotalaphdiag}a). The phase diagram shows
the clear correspondence between signatures in temperature and
field sweeps of the resistivity. The AF1 phase is fully suppressed
at $H^a_{c1}$ below 3~T. An uncertainty seems to remain from the
observation that the suppression becomes less rapid below 200~mK.
However, this slowdown of the suppression of $T_1$ with increasing
field is not observed in neutron scattering and
$H^a_{c1}=1.8\pm0.1$~T.\cite{fak05a}\ The discrepancy might have
been caused by the long relaxation to equilibrium of YbAgGe around
the first-order transition, which was observed in the
neutron-scattering experiment. The AF1 phase is followed by the
AF2 phase with its critical field $H^a_{c2}=3.0\pm0.1$~T, and by a
region III with unknown magnetic properties, which exists up to
$H^a_{c3}=4.9\pm0.1$~T. Figure~\ref{YbAgGetotalaphdiag}a shows how
the results of this study extend the phase diagram, obtained by
previous thermodynamic and transport measurements down to 0.4~K.
There is good agreement between our and previous data in the
overlap region from 0.4 to 0.75~K.\cite{bud04a,ume04a,bud04b}\ The
shape of signatures of the two transitions at $T_1$ and $T_3$ in
our and previous resistivity measurements is very similar as well.

\pgnfigurestwo{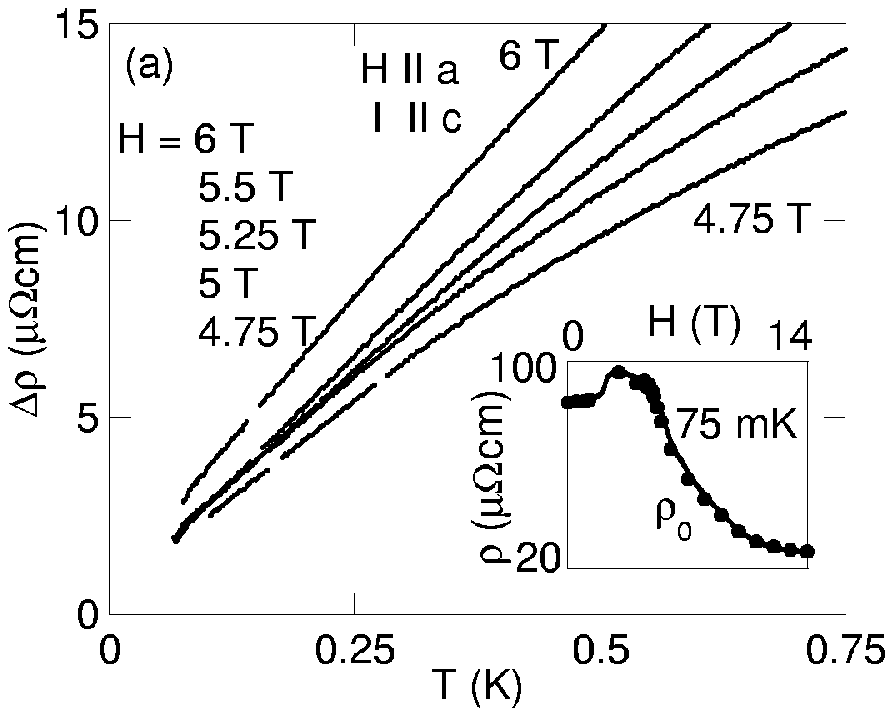}{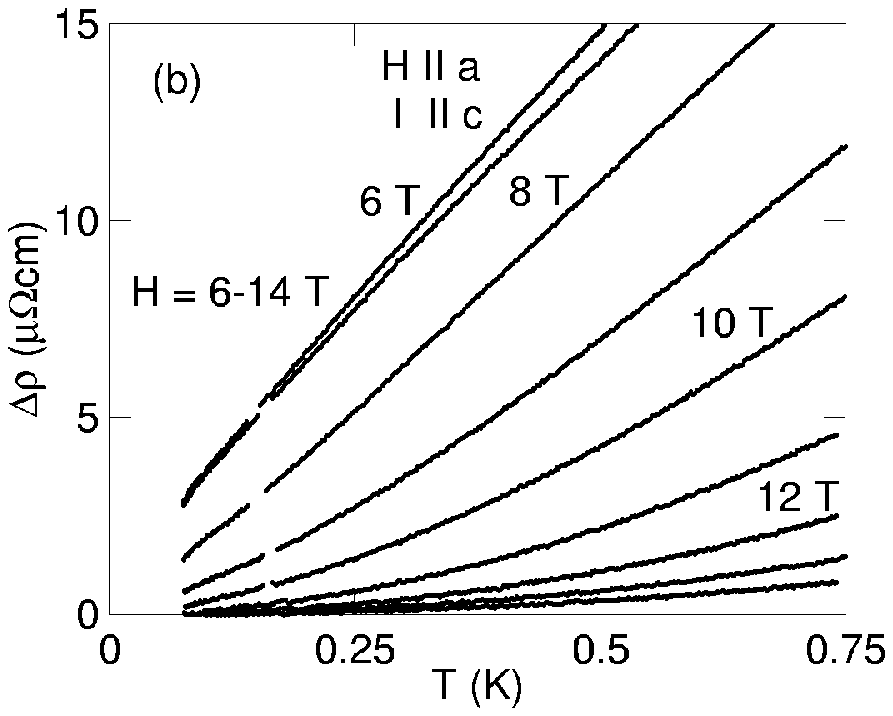}{Temperature
dependence of the resistivity for $H\|a$ and $I\|c$. The two
graphs together with Figure~\ref{YbAgGemcatswp1tr}b illustrate the
development of the low-temperature dependence of $\rho(T)$ with
$H\|a$. The inset in (a) shows the field dependence of
$\rho$(75~mK) and $\rho_0$ determined from power-law fits (filled
circles).}

\pgnfigurestwo{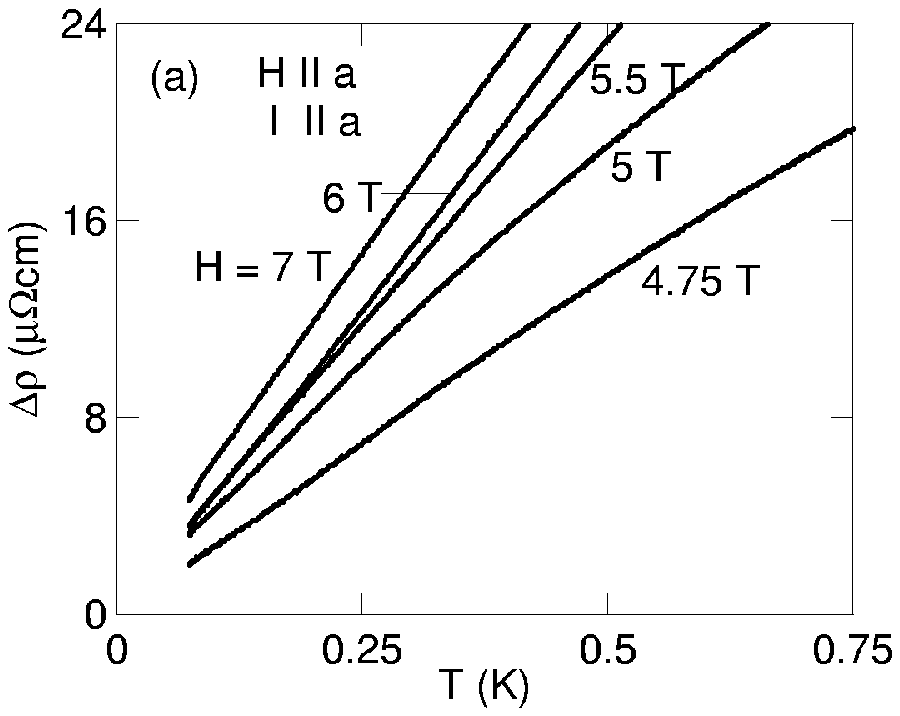}{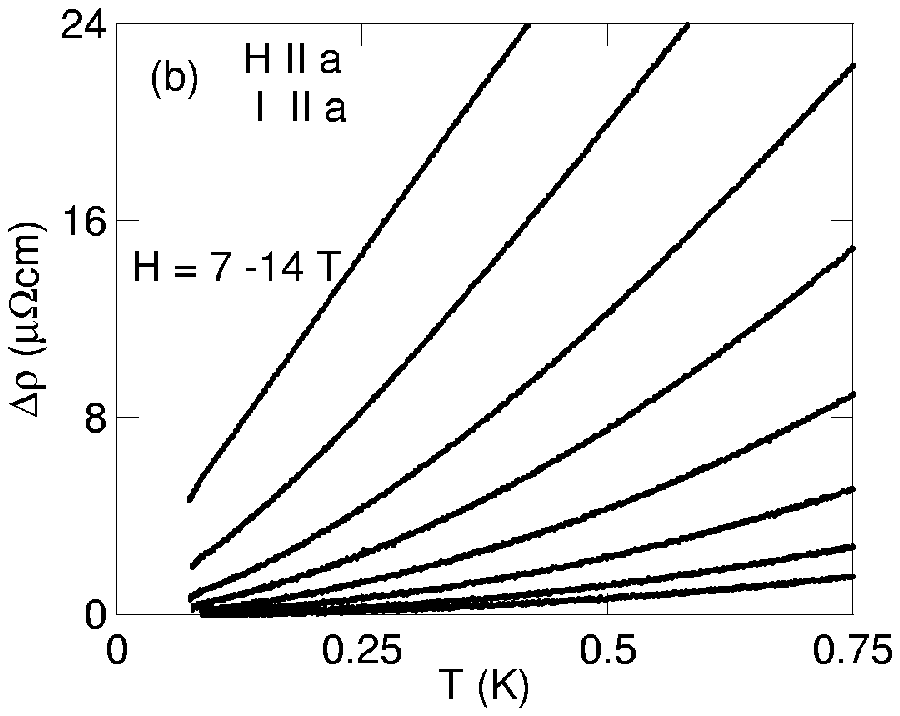}{Temperature
dependence of the resistivity for $H\|a$ and $I\|a$. The two
graphs illustrate the development of the low-temperature
dependence of $\rho(T)$ with $H\|a$.}

\begin{figure}\includegraphics[height=13cm,clip=true]
{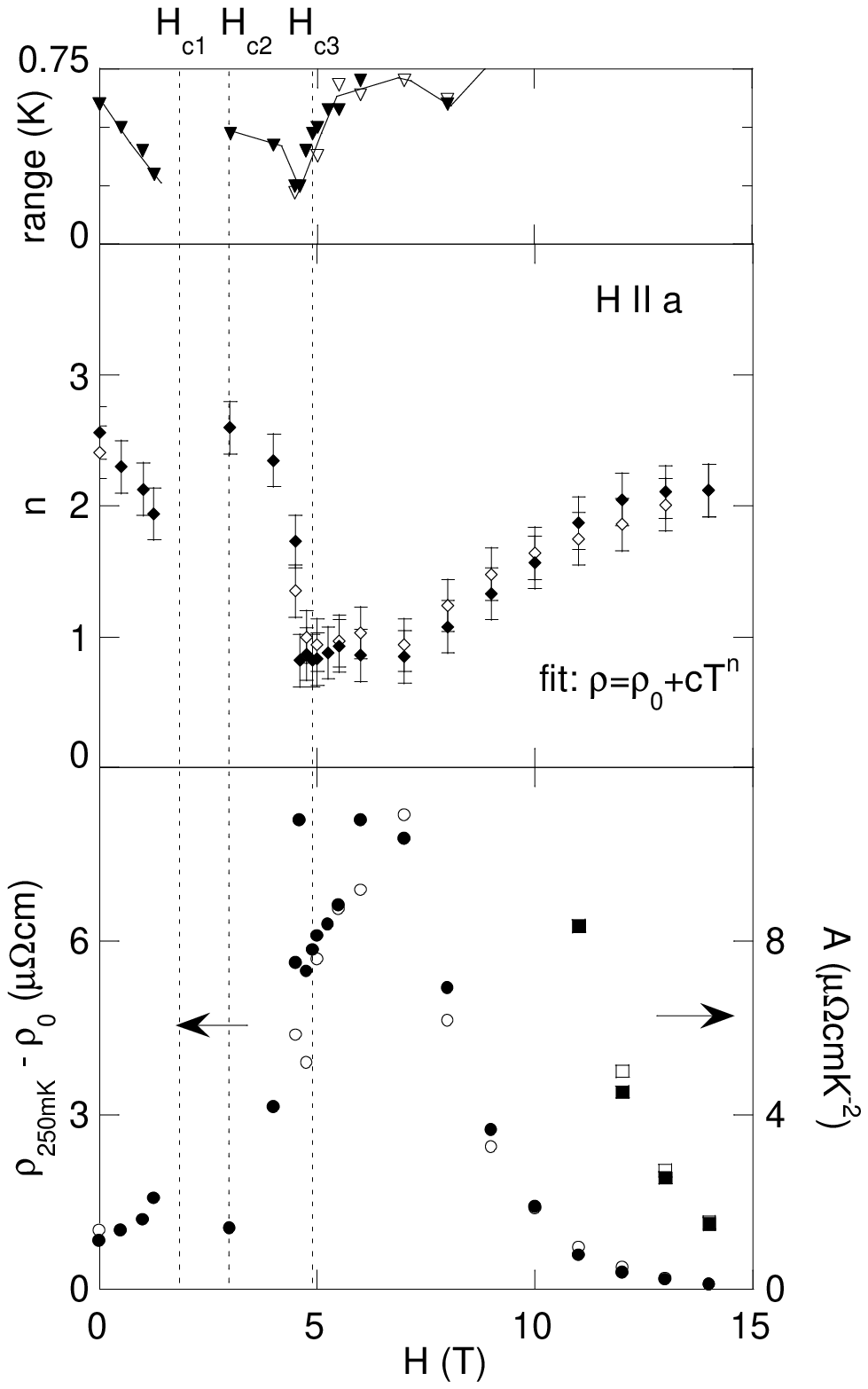}\caption{\label{YbAgGemalowTexp} Analysis of
the low-temperature resistivity for $H\|a$. Measurements with
current $I\| a$ (empty symbols) or $I\| c$ (filled symbols) give
similar results. The low-temperature exponent $n$ (diamonds)
results from single-power-law fits of $\rho(T)$. The range, over
which the single-power law holds, is indicated in the upper part
of the figure (down triangles). Close to $H^a_{c1}$ the jump in
the resistivity prevents the determination of $n$. Below
$H^a_{c3}$ $n$ is generally close to or above 2. Between
$H^a_{c3}$ and $H\|a=7$~T $n\approx 1$. At higher fields the
exponent is increasing to $n\approx 2$. The evolution of the $A$
coefficient (squares) can only be determined at high fields where
$n\approx 2$. $A$ is the result of the fit $\rho = \rho_0 + AT^2$
up to 750~mK. In general, the magnitude of the low-temperature
dependence is measured by the size of $\Delta\rho$(250~mK)
(circles). Note the maxima at $H\|a=4.75$~T and $H\|a=6.5\pm
0.5$~T. The results of $A$ and $\Delta\rho$(250~mK) for $I\|a$
have been scaled down by a factor 0.56. These graphs are the
result of analysing data shown in Figures~\ref{YbAgGemcatswp1tr},
\ref{YbAgGemcatswp2trb}, and \ref{YbAgGemaatswp2trb}.}\end{figure}

\pgnfigurestwo{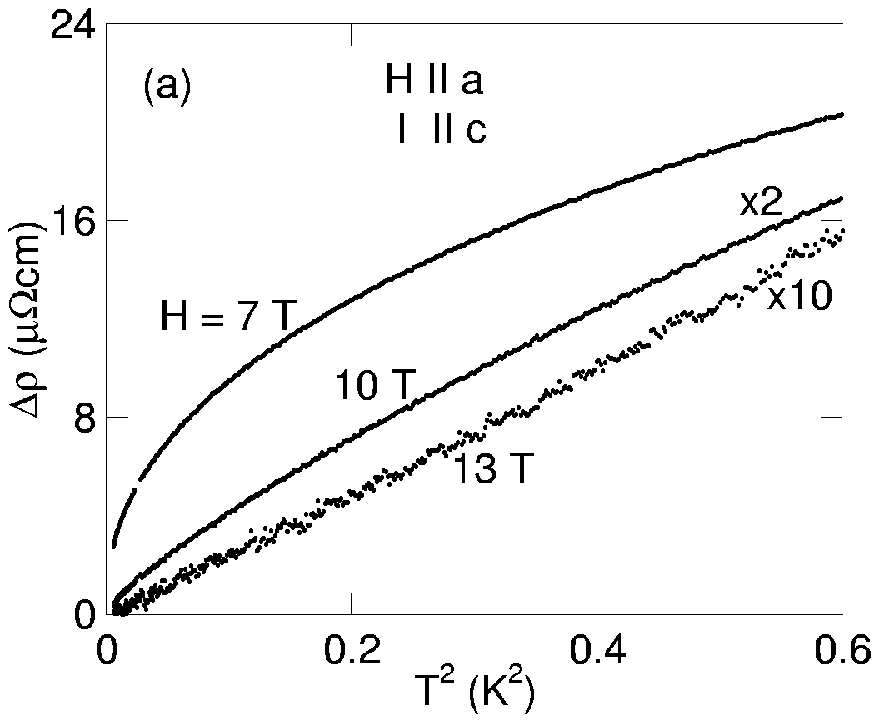}{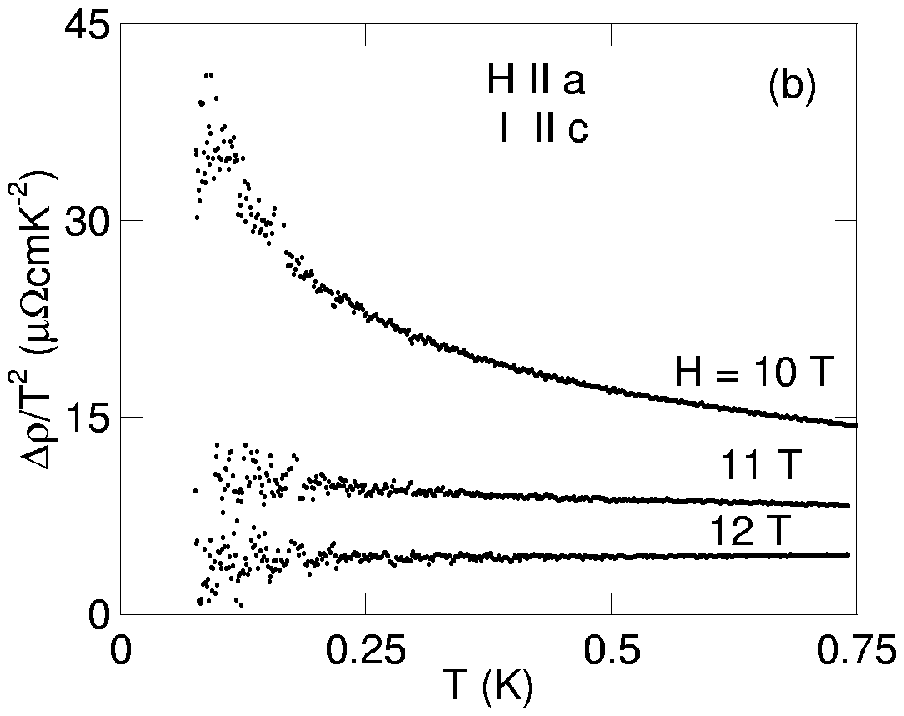}{Deviation of the
temperature dependence of the resistivity from quadratic behaviour
for $H\|a$. (a) At 7~T there is a clear deviation from a $T^2$
form of $\Delta\rho(T)$. The deviation is reduced towards higher
fields. The curves at high fields are shown on reduced scales to
allow better comparison of the qualitative temperature dependence.
(b) Above 10~T the resistivity becomes quadratic (horizontal line
of $\Delta\rho/T^2$ vs $T$). In this case, $\Delta\rho/T^2$
corresponds to the $A$ coefficient.}

We now give a more quantitative description of $\rho(T)$ of YbAgGe
when $H\|a$. In the inset of Figure~\ref{YbAgGemcatswp2trb}a the
field dependence of $\rho$(75~mK) and $\rho_0$ can be seen. On the
shown large resistivity scale both curves are similar.
$\rho_0(H\|a)$ drops strongly from low-field values of about
90~$\mu\Omega$cm to 25~$\mu\Omega$cm at 14~T. $\rho_0$ starts to
saturate at fields $H\|a$ of the order of 10~T, at which the
susceptibility is strongly reduced.\cite{bud04a}\

Figure~\ref{YbAgGemcatswp1tr}\ and \ref{YbAgGemcatswp2trb}\ for
$I\|c$ and Figure~\ref{YbAgGemaatswp2trb}\ for $I\|a$ demonstrate
the development with $H\|a$ of the magnitude of the
low-temperature dependence of $\Delta\rho$. The field dependence
of the temperature exponent $n$ obtained from power law fits is
shown in Figure~\ref{YbAgGemalowTexp}. At fields clearly below
$H^a_{c3}$ ($H\|a<4.5$~T) we find values for $n$ close to or above
the Fermi-liquid value 2. An approximately quadratic resistivity
is also observed at high fields $H\|a>10$~T. Around the highest
critical field $H^a_{c3}=4.9$~T we find $n=0.9\pm0.2$. It has to
be noted that $n\approx 1$ not only right at $H^a_{c3}$ but in an
extended field region up to 7~T, which agrees with the observation
of a linear $\Delta\rho$ in a wider field range above
0.4~K.\cite{bud04a}

Figure~\ref{YbAgGemalowTexp}\ also shows the development with
$H\|a$ of the magnitude of the temperature dependence of the
resistivity $\Delta\rho$ at low $T$. $\Delta\rho(T')$ at a low
temperature $T'$ is a measure of the amount of low-energy
excitations present in a material, since they can act as
scatterers of the conduction electrons. In a material like YbAgGe,
low-energy excitations at low temperatures include particle-hole
excitations or magnetic excitations (e.g. spin fluctuations);
phonons can be neglected. If $\Delta\rho(T')$ grows, e.g., due to
a magnetic-field change, this indicates a softening of some
low-energy excitations. Therefore, magnetic critical fluctuations
should lead to an increase of $\Delta\rho(T')$ at low $T'$. A
typical quantity to measure the strength of the low-temperature
dependence would be the $T^2$ coefficient $A$ of a quadratic fit
$\rho=\rho_0+AT^2$. For not too strong interactions Fermi liquid
theory holds and magnetic excitations can be reexpressed as
renormalised particle-hole excitations. However, for YbAgGe
$\Delta\rho\propto T^2$ only in a limited high-field range. From
$H\|a=14$~T to $H\|a=11$~T the value of $A$ rises by more than a
factor 5. For a comparison of the low-temperature dependence of
$\rho$ in the whole investigated field range we use the quantity
$\Delta\rho$(250~mK). Moreover, the field dependence is
qualitatively the same for $\Delta\rho(T)$ with, e.g., $T=150$~mK,
$T=250$~mK, or $T=350$~mK. $\Delta\rho$(250~mK) rises strongly
when approaching $H\|a=$6-7~T. A second peak in
$\Delta\rho$(250~mK) just below $H^a_{c3}$, which is very sharp,
might be an artefact caused by the feature at $T_3(H)$.
Figure~\ref{YbAgGemalowTexp} shows that the two peaks mark the
field range in which $n$ is minimal.

In Figure~\ref{YbAgGemcasqdev}\ we visualise the result of the
data analysis that $n<2$ in a considerable range of $H^a$.
$\Delta\rho(T^2)$ at $H\|a=7$~T deviates strongly from a straight
line, which would indicate a quadratic temperature dependence
(Figure~\ref{YbAgGemcasqdev}a). The more sensitive test of a
quadratic temperature dependence $\Delta\rho/T^2(T)$ shows a
considerable deviation up to $H\|a=10$~T
(Figure~\ref{YbAgGemcasqdev}b). A crossover from conventional
Fermi-liquid (FL) to non-Fermi-liquid resistivity (see
Figure~\ref{YbAgGetotalaphdiag}) sets in where $\Delta\rho(T^2)$
curves show deviations from straight lines. The error comes from
averaging the crossover temperatures for $I\| a$ and $I\| c$. Due
to the data-noise level, a determination of the resistivity
exponent $n$ is only possible above 150~mK. Our results confirm
indications for an unconventional resistivity in an extended field
range above $H^a_{c3}$ seen in previous resistivity measurements
down to 0.4~K.\cite{bud04a}

\subsection{$H\|c$}

\pgnfigure{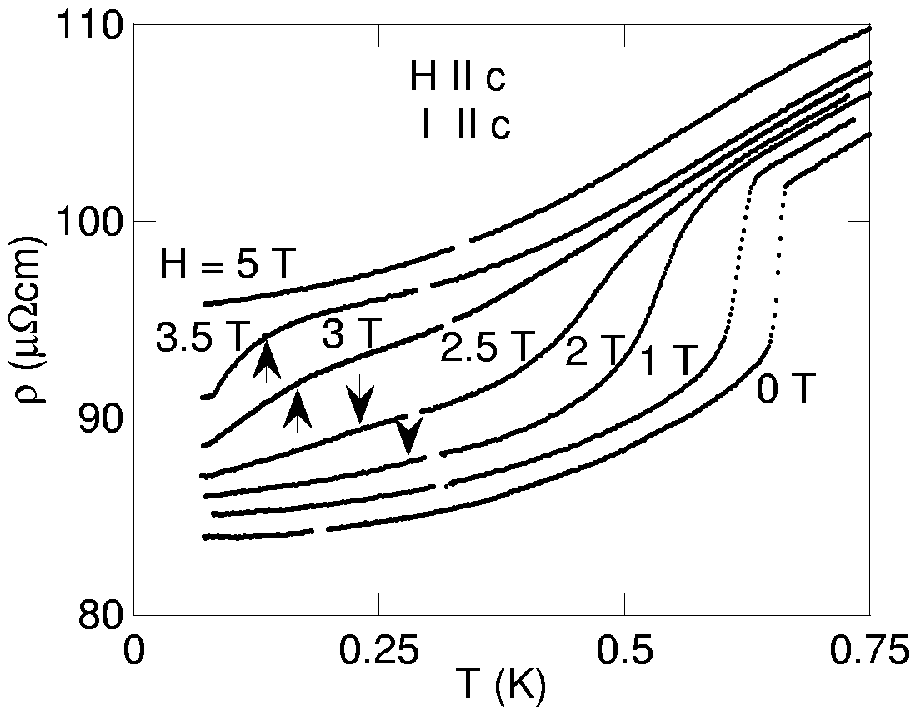}{Signatures of magnetic phase
transitions in $\rho(T)$ for $H\|c$. The first transition ($T_1$,
onset of a drop of $\rho$) is suppressed with field and fully
disappears at 5~T. The suppression does not happen continuously.
At 2~T and above, the jump-like drop gets washed out and a
shoulder ($T^*$, indicated by arrows) occurs.}

\pgnfigure{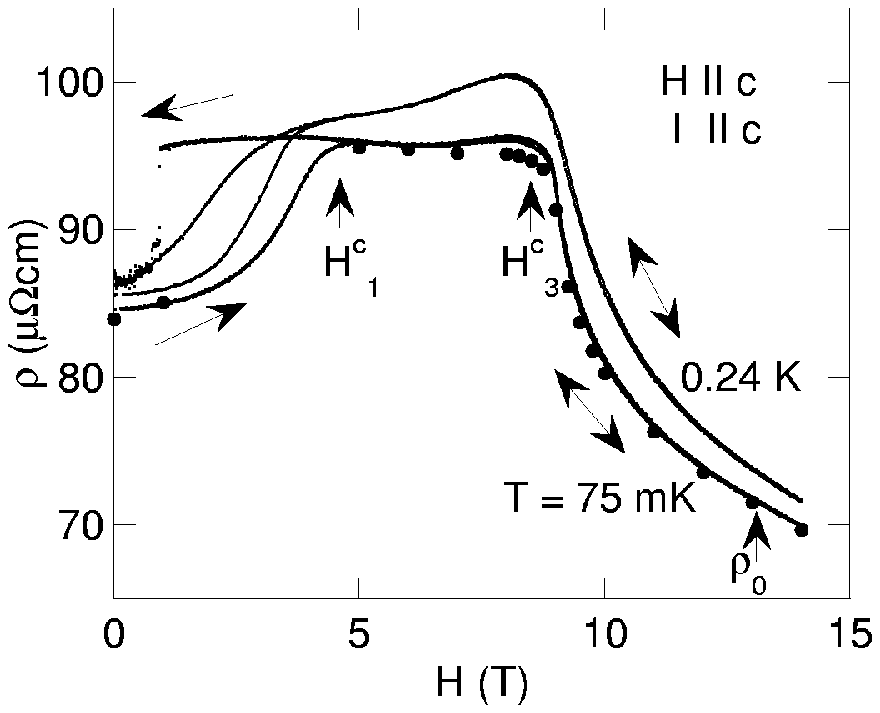}{Signatures of the two magnetic phase
transitions in $\rho(H\|c)$ at low temperatures. The large arrows
indicate the directions of the field sweeps. The sweeps show the
clear first order character of the first transition. Due to the
small temperature dependence of $\rho$ below 100~mK, the field
sweeps at 75~mK are very similar to $\rho_0(H\|c)$ determined from
power-law fits (filled circles). The difference between $\rho(H)$
at 75~mK and 240~mK illustrates the $H\|c$ dependence of the
strength of the low-temperature resistivity.}

Now the results for the resistivity of YbAgGe obtained for $H\|c$
will be described. We exclusively measured the c-axis resistivity.
Figure \ref{YbAgGemcctswp1tr} shows the suppression of the
first-order transition with increasing field. At $H\|c=1$~T it
still resembles a jump-like drop known from zero-field
measurements and measurements with $H\|a$. This jump is suppressed
to lower temperatures with increasing $H\|c$. However, from 2~T
onwards the jump becomes considerably washed out. At the same time
a shoulder appears at temperatures clearly below the position of
the washed out jump. With increasing $H\|c$ this shoulder becomes
more clearly visible but suppressed in temperature as well. At
$H\|c=5$~T the signature of the first transition can no longer be
detected above 70~mK. No clear features of the second-order
transition can be detected in $\rho(T)$ at higher fields. In
$\rho(H\|c)$, however, two transitions are visible: the transition
field of the first-order transition $H^c_1(T)$ is marked by the
onset of a drop in $\rho(H\|c)$ towards lower fields and the
transition field $H_3^c(T)$ at higher fields is marked by a
shoulder (Figure~\ref{YbAgGemcchswp}). This transition is denoted
as $H_3^c(T)$ and $T_3$ because of the similar features in
$\rho(H)$ and the low-temperature Hall resistivity\cite{bud04b}\
that are used to identify it (in analogy to the $H\|a$ data). The
hysteresis in $\rho(T)$ around $H_1^c$ is further evidence for the
first-order character of the first transition. The signatures for
phase transitions in $\rho(T,H\|c)$ are summarised in the
$H^c$-$T$ phase diagram (Figure~\ref{YbAgGetotalaphdiag}b). The
onset of the drop of $\rho(H\|c)$ towards lower fields seems to
correspond to the jump-like feature ($T_1$) in $\rho(T)$, marking
at low fields the transition from region I to III (which
corresponds to the transition between AF1 and AF2 order at zero
field). However, from $H\|c=2$~T onwards, at $T^*$ separating
region I and II, YbAgGe might undergo some transition or crossover
(indicated by the shoulder-like feature in $\Delta\rho(T)$). The
critical field of region I is $H^c_{c1}=4.4\pm0.1$~T. Above
$H^c_{c1}$, region III extends to $H^c_{c3}=8.8\pm0.1$~T. The
results of this study extend previous results from thermodynamic
and transport measurements down to 0.4~K (see discussion section)
and agree well with previous results in the overlap
region.\cite{bud04a}

\pgnfigurestwo{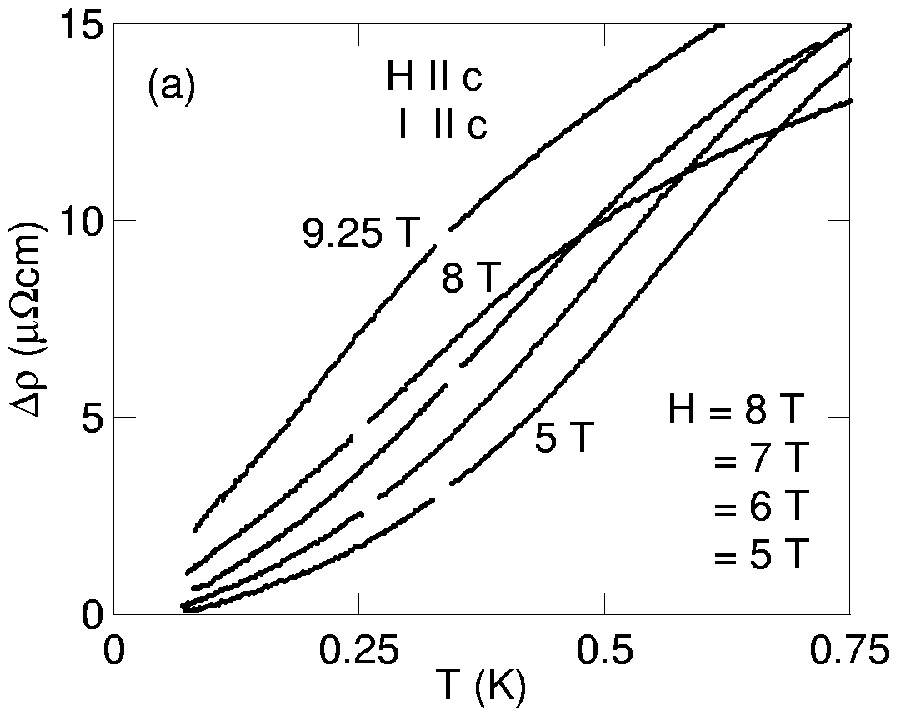}{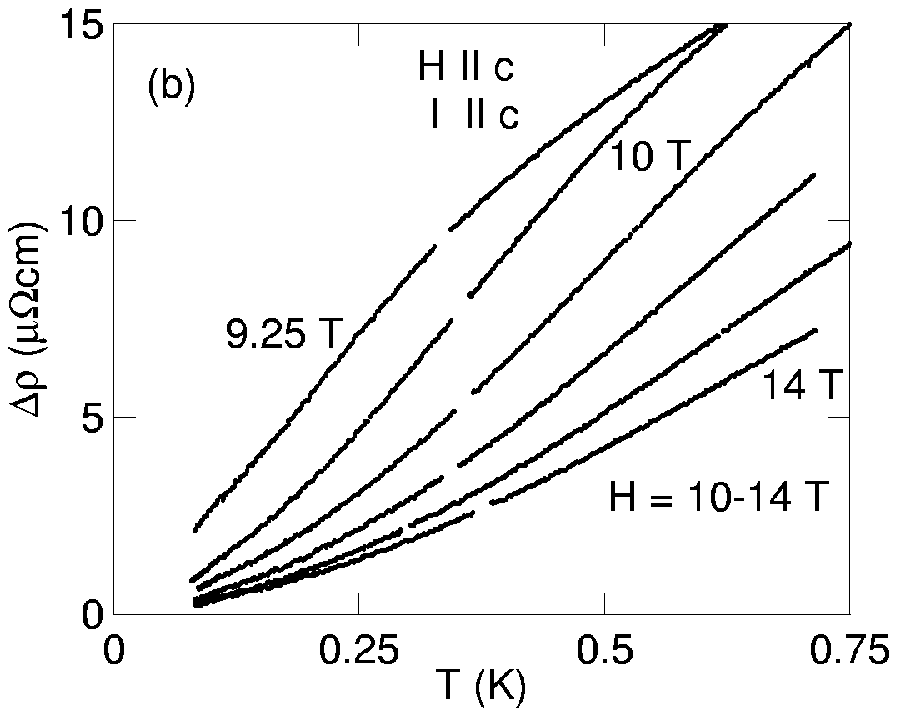}{Temperature
dependence of the resistivity for $H\|c$. The two graphs
illustrate the development with field of the low-temperature
dependence of $\rho(T)$, which is strongest at 9.25~T.}

\begin{figure}\includegraphics[height=13cm,clip=true]
{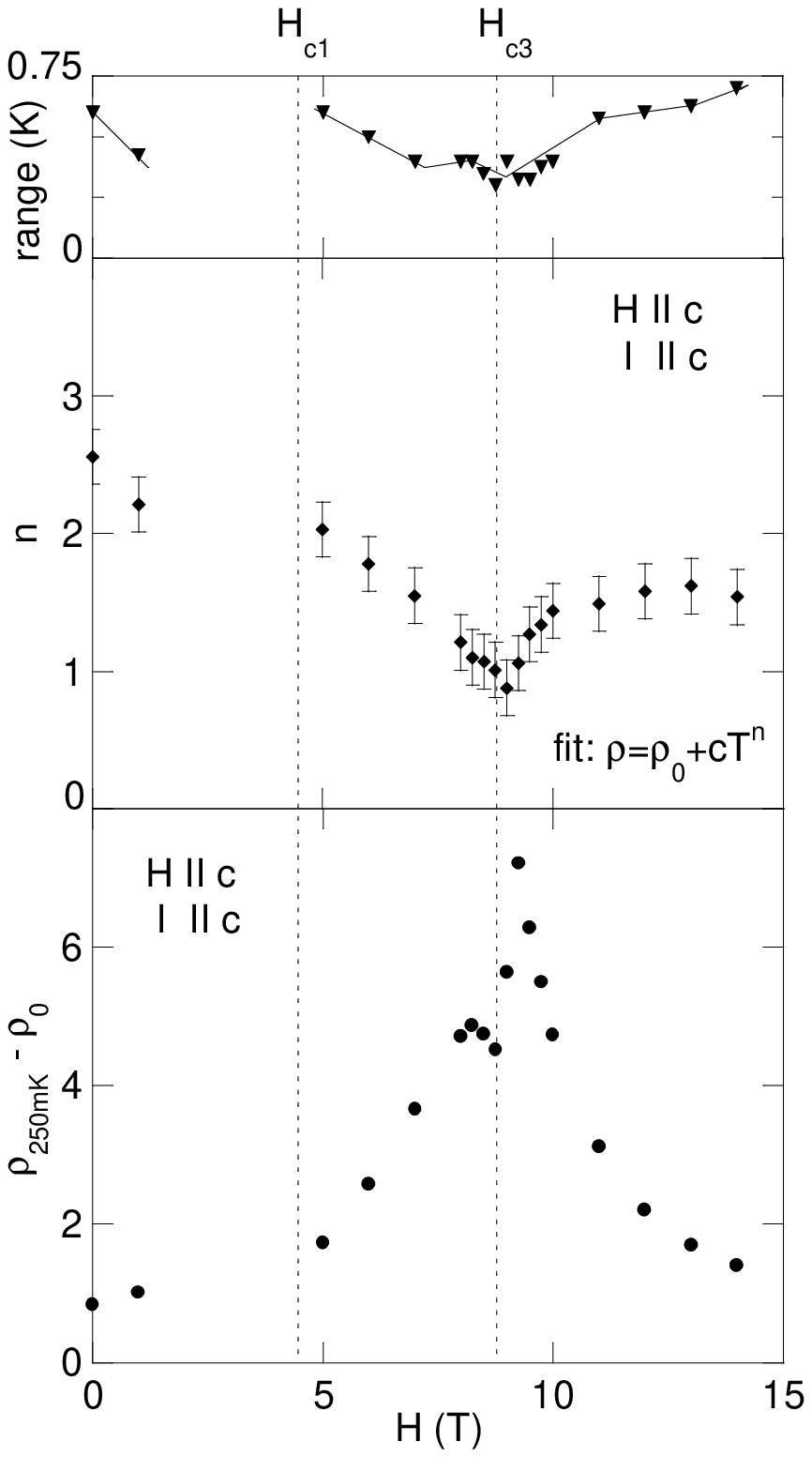}\caption{\label{YbAgGemcclowTexp} Analysis
of the low-temperature resistivity for $H\|c$. The low-temperature
exponent $n$ results from single-power-law fits of $\rho(T)$
(diamonds). The range, over which the single-power law holds, is
indicated in the upper part of the Figure (down triangles). Close
to $H^c_{c1}$ the shoulder in $\rho(T)$ does not allow a
determination of $n$. In the AF1 phase $x\ge 2$. At higher fields
$n<2$ in a considerable range with a minimum $n\approx 1$ close to
$H^c_{c3}$. The magnitude of the low-temperature dependence of
$\rho(T)$ is measured by the size of $\Delta\rho$(250~mK) and
reaches a maximum close to $H^c_{c3}$. These graphs are the result
of analyzing measurements shown in
Figure~\ref{YbAgGemcctswp2tra}.}\end{figure}

The $H\|c$ dependence of $\rho$(75~mK) and $\rho_0$
(Figure~\ref{YbAgGemcchswp}) are similar on the shown large
resistivity scale. A strong drop of $\rho_0$ towards high fields
is observed.

The $H\|c$ dependence of the low-temperature dependence of
$\Delta\rho$ can be seen in Figure~\ref{YbAgGemcctswp2tra}. The
magnitude (slope) of the low-temperature dependence of
$\Delta\rho$ is generally increasing at lower fields. At
$H\|c=9.25$~T it is reaching a maximum and then decreasing towards
high fields. To discuss the field dependence of $\Delta\rho$ more
quantitatively, $\rho$ has again been fitted with a general power
law of the form $\rho=\rho_0+cT^n$ and $n(H\|c)$ is shown in
Figure~\ref{YbAgGemcclowTexp}. For finite $H\|c<7$~T $n$ is close
to the Fermi-liquid value 2. Towards $H^c_{c3}$ the temperature
exponent is falling to $n\approx 1.0\pm0.2$ in an interval of 1~T
around the critical field. At higher fields $n>1$ up to 14~T.

Figure~\ref{YbAgGemcclowTexp}\ also shows the enhancement of the
magnitude of the low-temperature dependence of $\Delta\rho$
towards $H^c_{c3}$. The magnitude of the low-temperature
dependence of $\Delta\rho$ is again expressed by the quantity
$\Delta\rho$(250~mK). An enhancement of
$\Delta\rho(250~\mbox{mK})$ towards $H^c_{c3}$ is observed, which
corresponds to the field range where $n\approx 1$. The existence
of two peaks around $H^c_{c3}$ instead of one peak at $H^c_{c3}$
might be an artefact caused by the signal of the second
transition. Our results confirm indications for Non-Fermi-liquid
(NFL) resistivity in an extended field range close to and above
$H^c_{c3}$ seen in previous resistivity measurements down to
0.4~K.\cite{bud04a}

\section{Discussion}

First we interpret briefly the $H-T$ phase diagram obtained for
YbAgGe (Figure~\ref{YbAgGetotalaphdiag}). The transition at $T_1$
is first order, as is most evident from the large hysteresis in
$\rho(H\|c)$ (Figure~\ref{YbAgGemcchswp}). Temperature exponents
$n>2$ found in the spontaneously ordered phases
(Figures~\ref{YbAgGemalowTexp} and \ref{YbAgGemcclowTexp})
indicate scattering from spin waves. Spin waves are present, if
the characteristic temperature for the freezing out of spin waves,
which is determined by the band splitting in the magnetically
ordered phases, is small enough. Considering that at 1~K only a
small amount of entropy was detected in heat-capacity measurements
at zero field the ordered moment in the AF1 phase and the related
band splitting should be small and the presence of spin waves
likely. However, no spin-wave excitations have been found so far
in inelastic neutron experiments.\cite{fak04a}\ Around the other
lines in the phase diagram no hysteretic behaviour has been found.
Neutron scattering measurements suggest a suppression of the AF2
phase at $T_2$ via a second order transition.\cite{fak05a}\ It has
yet to be clarified, whether the features at $T^*$ and $T_3$
correspond to crossovers or phase transitions. Therefore, the
existence of QCPs at the corresponding critical fields is
possible.

\pgnfigurestwo{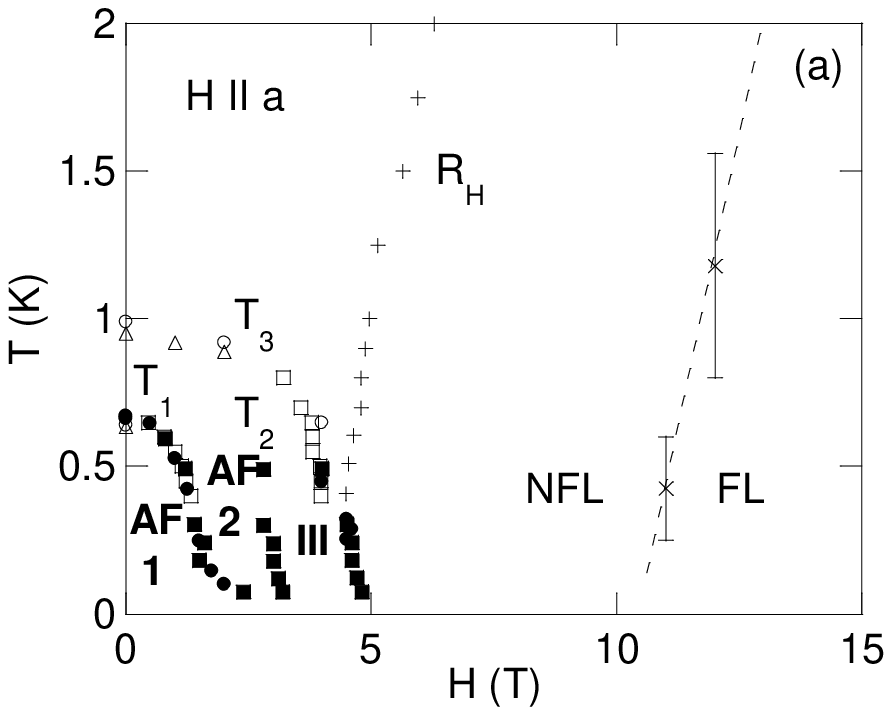}{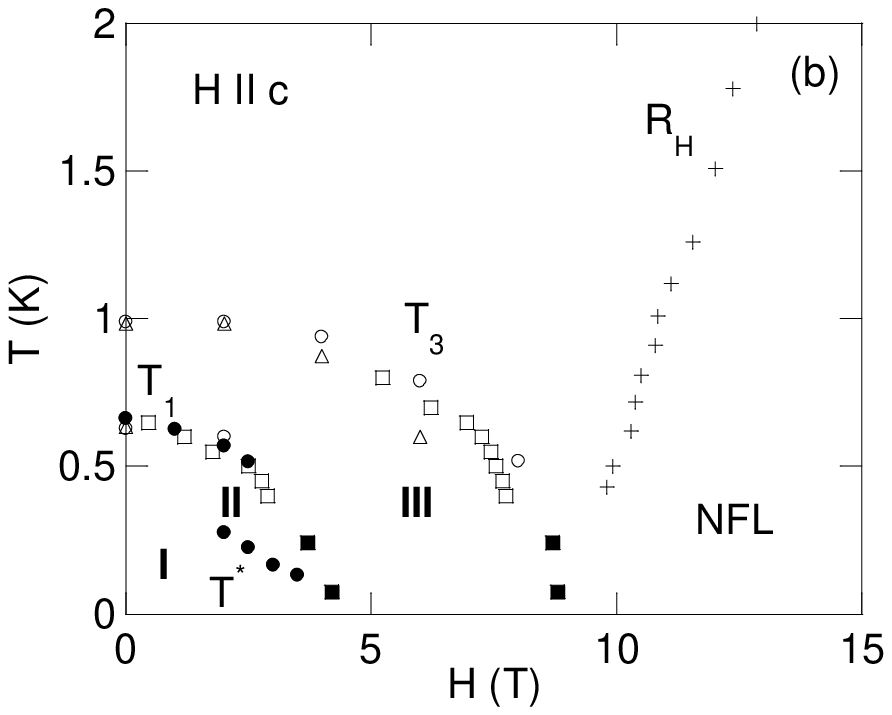}{Field-temperature
phase diagrams of YbAgGe. The magnetic phase boundaries are mapped
by combining results of this study (representative data sets shown
in Figures~\ref{YbAgGemcatswp1tr}, \ref{YbAgGemaahswptr},
\ref{YbAgGemcctswp1tr}, and \ref{YbAgGemcchswp}; filled symbols)
and a recent study (empty symbols).\cite{bud04a}\ Indications for
phase transitions have been seen in the resistivity (from
up-sweeps in $T$, circles), the magnetoresistance (from up-sweeps
in $H$, squares), and in the specific heat (triangles). Features
in the Hall resistivity (crosses, $R_H$) define a further line in
the phase diagrams.\cite{bud04b}\ (a) For $H\|a$ at lower fields
two antiferromagnetic phases (AF1 and AF2) are found, as well as
region III with yet unknown magnetic properties, and a high field
regime. The crossover in the high-field regime between
non-Fermi-liquid (NFL) and Fermi-liquid (FL) resistivity sets in
where $\Delta\rho(T^2)$ ceases to follow a straight line. This
phase diagram is the combined result of the in-plane and c-axis
resistivity. (b) For $H\|c$, the various order parameters are
unknown but region I and III correspond to AF1 and AF2 order at
zero field, respectively. At high fields an NFL regime is found.}

A main aim of the presented work was to study in detail the
development of unconventional material properties in approaching
the critical fields in YbAgGe. The results for the field and
temperature dependent resistivity $\rho(T,H)$ should serve to
address the question whether a magnetic field-induced QCP is
present in this material. Indeed, $\rho$ indicates the existence
of a field-induced QCP in YbAgGe in two ways when approaching the
vicinity of the highest critical field: (i) the temperature
dependence of resistivity measured by $\Delta\rho$(250~mK) is
strongly increased and (ii) the temperature exponent $n$ decreases
to 1, i.e. clearly dropping below the Fermi-liquid value.
Currently, detailed predictions for the resistivity in the
vicinity of a field-induced QCP are still lacking. However, the
unconventional features observed in the resistivity of YbAgGe
resemble qualitatively predictions for the resistivity of a
material which is tuned through a QCP at zero field. In the
spin-density wave scenario, the low-temperature exponent $n$ is
predicted to be below 2 at the critical point and can be as low as
1 depending on the dimensionality $D+z$ of the spin system, where
$D$ is the number of spatial dimensions and $z$ the dynamical
exponent.\cite{her76a,mil93a,mor85a,lon97a}\ Antiferromagnetic
fluctuations act only as scatterers of quasiparticles from the
so-called hot regions of the Fermi surface and $n=2$ for perfectly
pure samples. However in the limit $\Delta\rho(T)/\rho_0\ll 1$,
which is relevant for this measurement, effects from
short-circuiting of hot regions are
insignificant.\cite{hlu95a,ros99a,ros00a}. A linear resistance is
predicted for three-dimensional electrons coupled to
two-dimensional critical fluctuations.\cite{ros97a}\ Furthermore,
in the spin-density-wave scenario a strong increase of the general
temperature dependence of the resistivity is expected close to the
QCP due to scattering from a strongly increased amount of
fluctuations of softened magnetic modes. We add that $\Delta C/T$
of YbAgGe does not saturate in measurements down to
0.4~K.\cite{bud04a} This pattern of unconventional low-temperature
resistivity and heat capacity has been observed a number of times
in, e.g., heavy fermion systems or weakly magnetic d-metals when
approaching a zero-field QCP by changing the chemical composition
or applying pressure.

The observation of an NFL resistivity to low temperatures in a
wide field region is unusual even in the context of field-induced
quantum criticality. There are no indications for phase
transitions at the high-field side of $H^a_{c3}$ or $H^c_{c3}$ and
therefore the wide NFL range is not created by two nearby QCPs.
Local variations of the magnetic properties due to impurities or
defects might not be a good explanation of the wide NFL range
either, considering the sharpness of the first-order transition.
MnSi shows deviations from Fermi-liquid resistivity in a wide
pressure range\cite{the97a,pfl97a,pfl01b,doi03b}\ but the relevant
underlying physics close to a first-order transition to a
helimagnetic state appears to be quite different from the
mechanisms important in YbAgGe close to $H_{c3}$. However, we note
that the NFL range matches the field range, over which $\rho_0$
drops strongly and beyond which the magnetic susceptibility is
strongly reduced.\cite{bud04a}\ Such behaviour has been observed
in other rare-earth metals like CeAl$_2$ or CePb$_3$ (Refs.
\onlinecite{lap87a,wel87a}) and marks the regime of competition
between internal magnetic interactions and the Zeeman effect
caused by the external magnetic field. Given the large field
dependence of $\rho_0$ in the intermediate field regime, it will
be interesting to study the impurity dependence of the NFL range.

In YbAgGe, antiferromagnetic order at low magnetic fields suggests
the importance of antiferromagnetic interactions in this material.
The competition between internal antiferromagnetic interactions
and the Zeeman interaction is likely to be at the origin of
enhanced critical fluctuations. Currently, it is unclear by which
model the spectrum of the critical fluctuations can be best
described. Apart from the spin-density-wave scenario the local
character of the f-electrons might be important to categorise a
QCP in YbAgGe taking into account the low Kondo temperature
$T_K\approx 25$~K.\cite{mor04a,kat04a}\ In this respect YbAgGe
could turn out to be similar to CeCu$_{6-x}$Au$_x$ at its
zero-field QCP, or similar to YbRh$_2$Si$_2$ at its QCP near
zero-field. There unconventional transport and thermodynamic
properties are assigned to the strong localisation of the
f-electrons in these systems.\cite{sch00a,cus03a}\. We note the
formal similarity in the phase diagram of YbRh$_2$Si$_2$ and
YbAgGe that a feature in the hall resistivity defines a line which
originates from the critical field for the suppression of
magnetism.\cite{pas04a,bud04b}\ Finally, frustration effects in
the quasi-Kagome planes might be a relevant ingredient to the
fluctuation spectrum of YbAgGe. The fluctuation spectrum has to be
studied directly by in-field neutron scattering measurements and
in-field heat capacity measurements below 0.4~K to learn more
about the nature of the low-energy excitations in YbAgGe. The
impurity-level dependence of the size of the field range, where
YbAgGe shows non-Fermi-liquid resistivity, should also be
investigated.

\section{Conclusion}

The presented measurements of the field and temperature dependent
resistivity of single crystals of hexagonal YbAgGe up to 14~T
allow to extend the $H-T$ phase diagrams down to 70~mK. In
particular, for the field applied along the crystallographic
a-axis, the critical fields for the suppression of two
antiferromagnetic phases have been determined to be
$H^a_{c1}\approx 2$~T, $H^a_{c2}=3.0\pm0.1$~T. A further phase is
suppressed at $H^a_{c3}=4.9\pm0.1$~T. For the field applied along
the crystallographic c-axis, critical fields for the suppression
of phases, which show AF1 order and AF2 order at zero field, are
$H^c_{c1}=4.4\pm0.1$~T and $H^c_{c3}=8.8\pm0.1$~T, respectively.
In the low-field magnetically ordered phases the low-temperature
dependence of the resistivity is generally characterised by a
temperature exponent $n$ close to or above 2. However, close and
beyond the highest critical fields for both field directions
unconventional exponents $1\le n<1.5$ describe the low-temperature
resistivity, before Fermi-liquid behaviour is approached at high
fields. Since unconventional temperature exponents are accompanied
by a strong enhancement of the strength of the low-temperature
dependence of $\rho(T)$ our results indicate the existence of
field-induced quantum critical fluctuations in YbAgGe whose nature
has yet to be specified.


\begin{acknowledgments}

We thank B. F\aa k, Ch. R\"uegg, D. McMorrow, A. Huxley, M.
Zhitomirsky, H. Tsunetsugu, and M. Continentino for valuable
discussions on this topic. One of the authors (PGN) thanks P. Haen
for a seminar on magnetoresistance measurements. Ames Laboratory
is operated for the U.S. Department of Energy by Iowa State
University under Contract No. W-7405-ENG-82. This work was
supported by the Director for Energy Research, Office of Basic
Energy Sciences.

\end{acknowledgments}


\end{document}